\documentclass[aps, pra, lengthcheck,letter]{revtex4}
\usepackage{amsfonts, amssymb, amsmath, graphicx}

\usepackage{dcolumn}
\usepackage{bm}
\usepackage[hypertex]{hyperref}

\begin{document}

\title{Population imbalance and pairing in the BCS-BEC crossover of three-component ultracold fermions}

\author{Tomoki Ozawa}
\author{Gordon Baym}

\affiliation{
Department of Physics, University of Illinois at Urbana-Champaign, 1110 West Green Street, Urbana, Illinois 61801, USA
}%

\date{\today}

\def\del{\partial}
\def\p{\prime}
\def\simge{\mathrel{%
         \rlap{\raise 0.511ex \hbox{$>$}}{\lower 0.511ex \hbox{$\sim$}}}}
\def\simle{\mathrel{
         \rlap{\raise 0.511ex \hbox{$<$}}{\lower 0.511ex \hbox{$\sim$}}}}

\newcommand{\feynslash}[1]{{#1\kern-.5em /}}

\begin{abstract}

We investigate the phase diagram and the BCS-BEC crossover of a homogeneous three-component ultracold Fermi gas with a U(3) invariant attractive interaction.
We show that the system at sufficiently low temperatures exhibits population imbalance, as well as fermionic pairing.
We describe the crossover in this system,
connecting the weakly interacting BCS regime of the partially population-imbalanced fermion pairing state and 
the BEC limit with three weakly interacting species of molecules, including pairing fluctuations within a \textit{t}-matrix calculation of the particle self-energies.
\end{abstract}

\maketitle

\section{Introduction}
Multi-component ultracold atomic systems have recently been the focus of both experiment and theory, motivated in part by the prospect of simulating a wider range of many-body models, such as lattice SU(N) models~\cite{Honerkamp_PRL_2004, cazalilla_2009, gorshkov_2010,takahashi_kyoto_2010} and quantum chromodynamics (QCD) analogs~\cite{rapp_color_2007, rapp_trionic_2008, wilczek_quantum_2007, maeda_2009}, than is possible with single- or two-component systems.
The possibility of creating analogs of color superfluid states and the formation of hadronic states in multicomponent systems~\cite{rapp_color_2007, rapp_trionic_2008, wilczek_quantum_2007} is especially interesting since the regime of cold dense QCD matter is not directly achievable in current nuclear experiments  or in lattice QCD.

When three species of fermions weakly attract each other, two species form Cooper pairs and the third remains a Fermi liquid~\cite{modawi_leggett, bedaque_superfluid_2006, paananen_pairing_2006, paananen_coexistence_2007, zhai_superfluidity_2007, catelani_2008, he_superfluidity_2006, cherng_superfluidity_2007}. Which two species pair depends on anisotropies in the interactions and mass differences between different species. If there is no anisotropy, the Hamiltonian of the system possesses global U(3) symmetry with respect to rotation in species space, and the pairing breaks this symmetry. 
An important feature of the three-component fermion system is spontaneous population imbalance, first noted in the continuum in Ref.~\cite{he_superfluidity_2006} at $T=0$.
In addition, BCS superfluidity and population imbalance (magnetism), with two independent order parameters, can coexist, an
intrinsic feature of a multicomponent Fermi systems,
as shown by Cherng et al.~\cite{cherng_superfluidity_2007} in the weak-coupling BCS regime.

We consider here U(3) invariant three-component ultracold Fermi gases in three-dimensional free space with 
varying interaction, with a focus on spontaneous population imbalance and superfluidity
 at finite temperature at general interaction strength; we
study the phase diagram in general and the BCS-BEC crossover of the system, fixing only the total number of particles and allowing spontaneous population imbalance to occur.  With a fixed total number of particles, population imbalance is accompanied by spatial inhomogeneities, such as, for example, domain formation.
We first analyze the system at zero temperature in BCS mean field  to show that the fermion pairing gap and population imbalance both develop with increasing bare attractive interaction between the fermions.
Then we discuss nonzero temperature, starting from the BCS region where the scattering length is small and negative.
We calculate the population imbalance as well as the BCS transition temperature there as a function of interaction strength and temperature, to lowest order in the interaction.
The thermodynamic potential derived here agrees with previous calculations ~\cite{bedaque_superfluid_2006,paananen_pairing_2006,paananen_coexistence_2007,zhai_superfluidity_2007,catelani_2008}  when the chemical potentials of the three species are equal. 
We also derive the Ginzburg-Landau free energy as a function of the two order parameter--the pairing
gap and the population imbalanc--and discuss a possible analogy between dense QCD and three-component ultracold fermions.  We then turn to the
 BEC limit of three-component ultracold fermions, where the scattering length is small and positive, a regime described by  three different weakly interacting species of molecules made of different combinations of fermions.
We show that Bose condensation of the molecules is accompanied by population imbalance.
Finally, we discuss the BCS-BEC crossover connecting BCS and BEC limits, following the procedure of Nozi\`eres and Schmitt-Rink \cite{nozieres_schmitt_rink} to include pairing fluctuations (or non-condensed pairs), here in a
summation of ladder diagrams for the self-energies;  this calculation yields a transition temperature to the condensate phase that reduces to the BCS and BEC  limits.

Degenerate three-component gases have been experimentally realized using the three lowest hyperfine states of $^6$Li~\cite{ottenstein_collisional_2008, huckans_three-body_2009};
at high magnetic fields, well beyond unitarity, the scattering lengths between the three hyperfine states are negative and sufficiently close that the system is approximately U(3) invariant. 
In addition, ultracold gases of alkaline-earth-metal atoms possess good SU(N) invariance (with N up to 10)~\cite{cazalilla_2009, gorshkov_2010, takahashi_kyoto_2010},
and are good candidates to observe the physics discussed here.
Ytterbium has an SU(6) symmetry due to the nuclear spin; an SU(3) invariant mixture can be obtained by using only three spin components.
In 
$^6$Li as well as in  $^{171}$Yb and $^{173}$Yb, the temperatures currently achieved experimentally are around $T \simge 0.3 T_F$ ~\cite{ottenstein_collisional_2008,huckans_three-body_2009, takahashi_kyoto_2010}.
With a factor of $\sim 3$ decrease in temperature, phase separation due to the formation of population-imbalanced domains could be observed.

Around the unitarity point, $1/a = 0$, in a U(3) invariant system (where $a$ is the s-wave scattering length), three-body Efimov bound states can exist~\cite{efimov_energy_1970,braaten_universality_2006, floerchinger_functional_2009, moroz_efimov_2009, floerchinger_three-body_2009,luu_three-fermion_2007,naidon_possible_2009,braaten_three-body_2009}.
Efimov states have been experimentally observed in a trap through an increase of the particle loss rate, mediated by these states \cite{ottenstein_collisional_2008, huckans_three-body_2009,grimm_2006}.
In this paper, we analyze the system on time scales long enough to see the two-body interaction physics but short enough that Efimov states or three-body collisions can be neglected; such an intermediate thermalized regime can exist in a trap at  sufficiently low densities, since 
the two-body collision rate is proportional to the particle density squared whereas the three-body collision rate is proportional to the density cubed \footnote{Huckans et al.~ \cite{huckans_three-body_2009} argue that strong interactions with a long lifetime ($>0.1$s) can  in fact be achieved in a low-density gas.}.
As we show, the homogeneous state is unstable against the formation of inhomogeneous structures with population imbalance; population imbalance suppresses the formation of Efimov states, tending to stabilize the inhomogeneous three-component system.

\section{Three-component U(3) invariant fermions}

We consider a three-component fermion system in free space with equal masses and the same scattering length between different species.
We label  the three species by ``colors" in analogy with QCD, ``red (r)," ``green (g)," and ``blue (b)." 
At low temperature, the interaction is dominated by s-wave scattering, and the Hamiltonian is
\begin{align}
	\mathcal{H}' \equiv\mathcal{H} - \mu \mathcal{N}
	&=
	\sum_{\mathbf{k}, \alpha}
	\left( \frac{k^2}{2m} - \mu \right) \psi^\dagger_{\alpha, \mathbf{k}} \psi_{\alpha, \mathbf{k}}
	\notag \\
	&+
	\frac{U}{2V}\sum_{\alpha, \beta} \sum_{\mathbf{k}, \mathbf{k}^\prime, \mathbf{q}}
	\psi^\dagger_{\beta, \mathbf{k}^\prime - \mathbf{q}} \psi^\dagger_{\alpha, \mathbf{k}+\mathbf{q}} \psi_{\alpha, \mathbf{k}} \psi_{\beta, \mathbf{k}^\prime}, \label{hamiltonian}
\end{align}
where $\psi^\dagger_{\alpha, \mathbf{k}}$ is the creation operator of a particle with color $\alpha$ = r, g, b with momentum $\mathbf{k}$; $V$ is the volume, and we take $\hbar =1 $ throughout.  We assume an attractive bare contact interaction of strength $U < 0$.  Although we take a common chemical potential $\mu$ for all three species, the numbers of each species in the state of lowest free energy can be different as a consequence of interactions,
an effect that would be observable,  in an experiment that starts with equal numbers, as an inhomogeneous state. The Hamiltonian is invariant under global U(3) rotations of the species.

The attractive interaction leads to pairing of fermions at low temperature.  The pairing order
parameter is antisymmetric in color, and thus has the form
\begin{align}
	\Delta_\alpha (\mathbf{r}) \propto \epsilon_{\alpha \beta \gamma}\langle \psi_\beta (\mathbf{r}) \psi_\gamma (\mathbf{r}) \rangle.
\end{align}
Since under a global U(3) rotation,
\begin{align}
	\psi_{\alpha, \mathbf{k}}
	\to
	U_{\alpha \beta}
	\psi_{\beta, \mathbf{k}},
\end{align}
where $U_{\alpha \beta} \in$ U(3)
(we use the
 convention that repeated indices are summed over),
 $\Delta_\alpha$ transforms as
\begin{align}
	\Delta_\alpha (\mathbf{r}) \propto \frac{\epsilon_{\alpha \beta \gamma}}{2}\langle \psi_\beta (\mathbf{r}) \psi_\gamma (\mathbf{r}) \rangle
	\to
	(\det U) U_{\alpha \beta}^* \Delta_{\beta} (\mathbf{r}),
	\label{deltatrans}
\end{align}
where $\psi_\alpha (\mathbf{r})$ is the Fourier transform of $\psi_{\alpha, \mathbf{k}}$.

To prove Eq.~(\ref{deltatrans}),
we consider the operator $\hat{\Delta}_\alpha = \epsilon_{\alpha \beta \gamma} \psi_{\beta} \psi_{\gamma}$, whose expectation value is proportional to $\Delta_\alpha$.
The combination $\psi_\alpha \hat{\Delta}_\alpha$ transforms as
\begin{align}
	\psi^T \hat{\Delta} &\equiv \psi_\alpha \hat{\Delta}_\alpha
	=
	\epsilon_{\alpha \beta \gamma} \psi_\alpha \psi_{\beta} \psi_{\gamma}
	\notag \\
	&\to
	\epsilon_{\alpha \beta \gamma} U_{\alpha \zeta}U_{\beta \eta} U_{\gamma \xi} \psi_\zeta \psi_{\eta} \psi_{\xi}
	\notag \\
	&=
	\det U \epsilon_{\zeta \eta \xi} \psi_\zeta \psi_{\eta} \psi_{\xi}
	=
	\det U \epsilon_{\alpha \beta \gamma} \psi_\alpha \psi_{\beta} \psi_{\gamma}
	\notag \\
	&=
	\det U \psi^T \hat{\Delta}.
\end{align}
On the other hand, $\psi^T \to \psi^T U^{T}$.
Therefore, $\hat{\Delta} \to \det U (U^{T})^{-1} = \det U U^*$.

As a consequence of the transformation ~(\ref{deltatrans}), we can -- when the order parameter is independent of position --  always choose appropriate axes of colors to transform the pairing order parameter into the form $\vec{\Delta} = (0, 0, \Delta)$, that is, by taking appropriate linear combinations of the species, we find that only two colors are paired and one is left unpaired.
By applying a Bogoliubov-Valatin transformation, we can see that there are two gapped fermionic excitations  corresponding to the quasiparticles of the paired fermions, and one ungapped excitation
 due to the unpaired fermions.
In the following, we assume, without loss of generality, that the red and green particles are paired and the blue are not paired.

\section{BCS Mean Field at $T = 0$ }
In this section, we consider the ground state of the system within mean-field BCS theory.
We describe the pairing between r and g particles and unpaired b particles with the BCS-like ansatz,
\begin{align}
	|\Psi\rangle
	=
	\prod_{\mathbf{k}}
	\left(
	u_{\mathbf{k}} + v_{\mathbf{k}} \psi^\dagger_{r, \mathbf{k}} \psi^\dagger_{g, -\mathbf{k}}
	\right)
	\prod_{|\mathbf{k}| \le k_F^b} \psi^\dagger_{b, \mathbf{k}} |\mathrm{vac}\rangle,
\end{align}
where $|u_\mathbf{k}|^2 + |v_\mathbf{k}|^2 = 1$ and $k_F^{b}$ is the b Fermi momentum.
The parameters $u_\mathbf{k}$ and $v_\mathbf{k}$ are determined by minimizing $\langle \Psi | \mathcal{H} - \mu \mathcal{N} | \Psi \rangle$ at fixed $\mu$.
Following the standard procedure, we obtain
\begin{align}
	u_\mathbf{k}^2
	=
	\frac{1}{2}
	\left(
	1
	+
	\frac{\xi_\mathbf{k}}{\sqrt{\xi_\mathbf{k}^2 + \Delta^2}}
	\right),
	\hspace{0.1cm}
	v_\mathbf{k}^2
	= 
	\frac{1}{2}
	\left(
	1
	-
	\frac{\xi_\mathbf{k}}{\sqrt{\xi_\mathbf{k}^2 + \Delta^2}}
	\right),
\end{align}
where $\xi_\mathbf{k} = k^2 / 2m - \mu$ and the gap $\Delta = -(U/V)\sum_\mathbf{k} u_\mathbf{k} v_\mathbf{k}$ is determined by 
\begin{align}
	\Delta
	=
	-\frac{U}{V}\sum_\mathbf{k}\frac{1}{2}\frac{\Delta}{\sqrt{\xi_\mathbf{k}^2 + \Delta^2}}.
\end{align}
We use the relation of the bare coupling $U$ and the scattering length $a$~\cite{galitskii,leggettbcsbec},
\begin{align}
	\frac{1}{U} = \frac{m}{4\pi a} - \frac{1}{V}\sum_{\mathbf{k}}\frac{m}{ k^2}, \label{renormalizeu}
\end{align}
to rewrite the gap equation for $\Delta \neq 0$ in terms of $a$ as
\begin{align}
	\frac{m}{4\pi a}
	=
	\frac{1}{V}\sum_\mathbf{k}\left( \frac{m}{ k^2} - \frac{1}{2}\frac{1}{\sqrt{\xi_\mathbf{k}^2 + \Delta^2}}\right).\label{gapt0}
\end{align}
The chemical potential is determined by fixing the total number of particles $N$:
\begin{align}
	N
	&=
	\langle \Psi | \sum_{\alpha, \mathbf{k}} \psi^\dagger_{\alpha, \mathbf{k}} \psi_{\alpha, \mathbf{k}} | \Psi \rangle
	\notag \\
	&=
	\sum_{\mathbf{k}}\left(1-\frac{\xi_\mathbf{k}}{\sqrt{\xi_\mathbf{k}^2 + |\Delta|^2}}\right) + V\frac{(k_F^{b})^3}{6\pi^2}. \label{numbert0}
\end{align}
The same gap and number equations were derived in Ref.~\cite{he_superfluidity_2006} using path-integral techniques.
We solve the gap equation (\ref{gapt0}) and the number equation (\ref{numbert0}) simultaneously to
calculate the pairing gap and the number imbalance in terms of the scattering length.

\begin{figure}[htbp]
\begin{center}
\includegraphics[width=9cm,keepaspectratio=true]{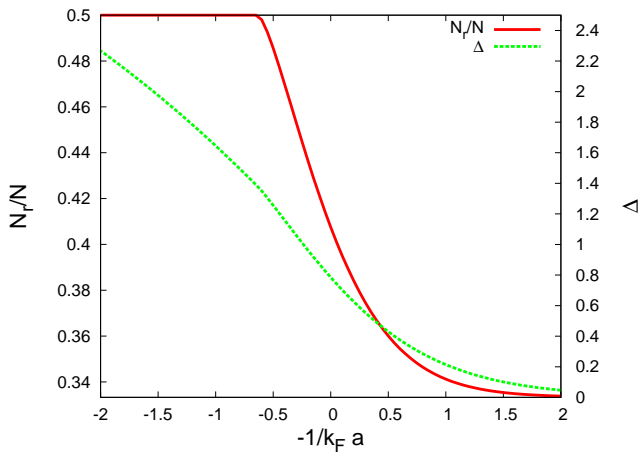}
\end{center}
\caption{(Color online) The number of red particles divided by the number of total particles $N_r/N$ and the pairing gap $\Delta$, in units of $\epsilon_F$, at zero temperature, vs. $-1/k_F a$. The solid line shows $N_r/N$ (left vertical axis) and the dotted line $\Delta$ (right vertical axis).}
\label{t0mean_gap_nr}
\end{figure}

In Fig.~\ref{t0mean_gap_nr}, we plot the pairing gap $\Delta$, measured in units of $\epsilon_F =  k_F^2 / 2m$, and the number of $r$ particles $N_r$ divided by the total number of particles $N$, against $-1/k_F a$, where $k_F = (6\pi^2 N/3V)^{1/3}$.
The right side of the figure corresponds to the weak-coupling regime (BCS region);
the bare coupling becomes stronger toward the left side (BEC region) of the figure.
As we see, $|\Delta|$ and the fraction of red particles, $N_r / N$, increase with stronger interaction.
The $N_r/N$ axis ranges from $1/3$ to $1/2$; when $N_r/N = 1/3$,  all three species are equally populated, but for
 $N_r/N = 1/2$,  only r and g particles are present. 
In general, $N_r/N$ is greater than 1/3 in the interacting system, and it approaches $1/2$ as the interaction becomes stronger.
Thus the ground state of the interacting system always exhibits population imbalance, or magnetization (in analogy with a spin system).
The magnetization arises physically through the gain of pairing energy when there are more particles in r and g states, and as remarked earlier, it would reveal itself in experiment as an inhomogeneous distribution of particle numbers.

With this basic picture in mind, we turn now to nonzero temperature.

\section{BCS region}

In the BCS region, where the scattering length $a$ is negative and small, perturbation theory in terms of the scattering length well describes the system.
We first derive the phase diagram in this region, and then we derive the corresponding Ginzburg-Landau free energy. 

\subsection{Mean-field phase diagram}

\begin{figure}[htbp]
\begin{center}
\includegraphics[width=8cm,keepaspectratio=true]{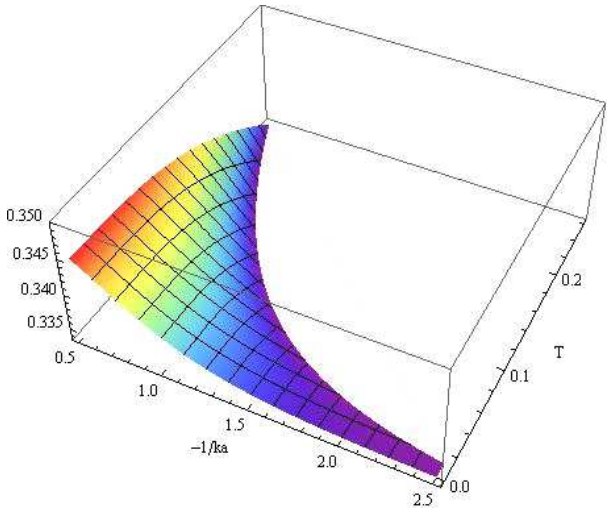}
\end{center}
\caption{(Color online) Phase diagram of the BCS region: $N_r/N$ vs. $-1/k_F a$ and temperature, in units of $\epsilon_F$. The
$z$-axis ranges from $1/3$ to $0.35$. The intersection of the surface and the bottom plane toward higher $T$ is the transition line between the ordered and normal phases.}
\label{tc_withHartree}
\end{figure}

The mean-field Hamiltonian $\mathcal{H}_M$ is
\begin{align}
	\mathcal{H}_M &- \mu \mathcal{N}
	\notag \\
	&=
	\sum_{\mathbf{k},\alpha} \left( \xi_\mathbf{k} + \frac{U_H}{V}(N-N_\alpha) \right) \psi^\dagger_{\alpha, \mathbf{k}} \psi_{\alpha, \mathbf{k}}
	\notag \\
	&\hspace{.1cm}
	-\Delta^* \sum_{\mathbf{k}}\psi_{r, \mathbf{k}} \psi_{g, -\mathbf{k}}
	-\Delta \sum_\mathbf{k} \psi^\dagger_{g, -\mathbf{k}} \psi^\dagger_{r, \mathbf{k}}
	\notag \\
	&\hspace{.2cm}-\frac{V}{U}|\Delta|^2
	-\frac{U_H}{V}\left(N_r N_g + N_g N_b + N_b N_r \right),
\end{align}
where 
\begin{align}
	\Delta
	=
	-\frac{U}{V}\sum_\mathbf{k} \langle \psi_{r, \mathbf{k}} \psi_{g, -\mathbf{k}} \rangle.
\end{align}
As was done earlier, we assume equal numbers of red and green particles, $N_r = N_g$.
Also, we now include the Hartree energy,  $U_H = 4\pi  a / m$.
Defining 
\begin{align}
	\xi_{r, \mathbf{k}}
	&=
	\xi_\mathbf{k} + \frac{U_H}{V} (N_r + N_b),
	\\
	\xi_{b, \mathbf{k}}
	&=
		\xi_\mathbf{k} + \frac{U_H}{V} 2N_r,
\end{align}
we rewrite the mean-field Hamiltonian as
\begin{align}
	\mathcal{H}_M &- \mu \mathcal{N}
	\notag \\
	&=
	\sum_{\mathbf{k}} \xi_{r, \mathbf{k}}
	\left(
	\psi^\dagger_{r, \mathbf{k}} \psi_{r, \mathbf{k}}
	+
	\psi^\dagger_{g, \mathbf{k}} \psi_{g, \mathbf{k}}
	\right)
	+
	\sum_{\mathbf{k}} \xi_{b, \mathbf{k}} \psi^\dagger_{b, \mathbf{k}} \psi_{b, \mathbf{k}}
	\notag \\
	&\hspace{.1cm}
	-\Delta^* \sum_{\mathbf{k}}\psi_{r, \mathbf{k}} \psi_{g, -\mathbf{k}}
	-\Delta \sum_\mathbf{k} \psi^\dagger_{g, -\mathbf{k}} \psi^\dagger_{r, \mathbf{k}}
	\notag \\
	&\hspace{.1cm}-\frac{V}{U}|\Delta|^2
	-\frac{U_H}{V}\left(N_r^2 + 2N_r N_b \right),
\end{align}
which is essentially the BCS mean-field Hamiltonian for paired red and green particles plus normal blue particles.
Diagonalizing by a Bogoliubov-Valatin transformation, we find
the thermodynamic potential 
\begin{align}
	&\Omega(T,\mu)
	=
	-\frac{2}{\beta}
	\sum_\mathbf{k}
	\ln \left[ 1 + e^{-\beta \epsilon_k} \right]
	-
	\frac{1}{\beta}
	\sum_\mathbf{k}
	\ln
	\left[
	1
	+
	e^{-\beta \xi_{b, \mathbf{k}}}
	\right]
	\notag \\
	&\hspace{.1cm}
	-
	\sum_\mathbf{k}
	\left(
	\varepsilon_k
	-
	\xi_{r, \mathbf{k}}
	\right)
	-\frac{V}{U}|\Delta|^2 - \frac{U_H}{V}\left(N_r^2 + 2N_r N_b \right),
\end{align}
where $\varepsilon_k \equiv \sqrt{\xi_{r, \mathbf{k}}^2 + |\Delta|^2}$.
The condition $\del \Omega / \del |\Delta|^2 = 0$ gives the gap equation
\begin{align}
	\frac1V\sum_\mathbf{k} \dfrac{1 - 2f\left( \varepsilon_k \right)}{2\varepsilon_k}
	=-\frac{1}{U}, \label{gapeqnonzerot}
\end{align}
where $f(x) = 1/(e^{\beta x} + 1)$ is the Fermi distribution function.
Again, $\mu$ is determined by the number equations
\begin{align}
	N_r
	&=
	\sum_\mathbf{k}
	\frac{1}{2}
	\left(
	1
	-
	\xi_{r, \mathbf{k}}\frac{\tanh \beta \varepsilon_k  / 2}{\varepsilon_k}\right),
	\\
	N_b
	&=
	\sum_\mathbf{k}
	f(\xi_{b, \mathbf{k}})
\end{align}
with
\begin{align}
	N = 2N_r + N_b. \label{numberbcs}
\end{align}
Numerically solving  the gap equation (\ref{gapeqnonzerot}) with the number equation (\ref{numberbcs}), we obtain
the gap and number imbalance at given temperature and scattering length, shown in
Fig. \ref{tc_withHartree}.
The figure plots $N_r / N$ as a function of $-1/k_F a$ and $T$.  The normal phase is the unshaded region at higher $T$;
here $\Delta = 0$ and $N_r / N = 1/3$.
In the shaded region, $\Delta \neq 0$ and $N_r / N >1/3$, a small number imbalance.
We show in the next subsection using the Ginzburg-Landau free energy that $\Delta \neq 0$ implies $N_r / N >1/3$ and vice versa.
 To extend the theory to the unitarity and BEC regimes, we take pair fluctuations into account
\footnote{We have so far assumed that the blue particles do not pair. However, the blue particles feel an effective attractive interaction with each other mediated by the existence of the red and green particles \cite{martikainen_2009}, which can
lead to p-wave pairing state of the blue particles.
However, as shown by Kagan and Chubukov \cite{kagan_1988}, the transition temperature to such p-wave pairing is 
too low ($T_c \sim 10^{-7}T_F$) to be observed in experiment, and we ignore it here.},
 in Sec. VI.

In the next subsection, we derive the Ginzburg-Landau free energy of the system in the BCS regime,
and derive the relations between the pairing gap and the number imbalance.

\subsection{Ginzburg-Landau free energy}

The interplay between pairing and number imbalance is most easily seen from the Ginzburg-Landau free energy, the expansion of the free energy in terms of the corresponding order parameters around the transition temperature.   We define the order parameter for number imbalance,  $\phi$, by
\begin{align}
	\phi
	=
	\frac{N_r}{V} - \frac{N}{3V},
\end{align}
Fixing the total number of particles $N = 2N_r + N_b$, we have equivalently
\begin{align}
\phi = -\frac12\left(\frac{N_b}{V} - \frac{N}{3V}\right).
\end{align}

To derive the Ginzburg-Landau free energy it is convenient (in the derivation only) to let the chemical potential $\mu_b$ for b be different from the chemical potential $\mu_r$ for r and g.
The thermodynamic potential $\Omega (T, \mu_r, \mu_b)$ can be derived as in the previous subsection.
The Helmholtz free energy is then
\begin{align}
	F(\Delta, \phi)
	&=
	\Omega + 2 \mu_r N_r+ \mu_b N_b,
\end{align}
in terms of which the Ginzburg-Landau (GL) free-energy density can be obtained by expanding
\begin{align}
	&\mathcal{F}_{\mathrm{GL}}(\Delta, \phi)
	\equiv
	\frac1V\left(F(\Delta, \phi) -F(0, 0)\right).
\end{align}
We  define
\begin{align}
	\tilde{\xi}_{r, \mathbf{k}} &= \frac{k^2}{2m} - \mu_r + \frac{U_H}{V} (N_r + N_b) 
	\\
	\tilde{\xi}_{b, \mathbf{k}} &= \frac{k^2}{2m} - \mu_b + \frac{U_H}{V} 2N_r,
\end{align}
and $\tilde{\varepsilon}_k \equiv \sqrt{\tilde{\xi}_{r, \mathbf{k}}^2 + \Delta^2}$,
and the chemical potential of the normal phase $\mu_0$ implicitly through
\begin{align}
	\frac{N}{3}
	=
	\sum_\mathbf{k} \frac{1}{e^{\beta \xi^0_{\mathbf{k}}} + 1},
\end{align}
where $\xi_\mathbf{k}^0 = k^2/2m - \mu_0 + 2U_H N/3V$.
In terms of these quantities, the Ginzburg-Landau free-energy density is
\begin{align}
	&\mathcal{F}_{\mathrm{GL}}(\Delta, \phi)
	\notag \\
	&\hspace{0.1cm}=
	-\frac{2}{\beta V}
	\sum_\mathbf{k}
	\ln
	\left[ 1 + e^{-\beta \tilde{\varepsilon}_k} \right]
	-
	\frac{1}{\beta V}
	\sum_\mathbf{k}
	\ln
	\left[ 1 + e^{-\beta \tilde{\xi}_{b, \mathbf{k}}} \right]
	\notag \\
	&\hspace{0.1cm}
	+
	\frac{3}{\beta V}\sum_\mathbf{k}
	\ln
	\left[ 1 + e^{-\beta \tilde{\xi}_{\mathbf{k}}^0} \right]
	-
	\frac{1}{V}
	\sum_\mathbf{k}
	\left(
	\tilde{\varepsilon}_k
	-
	\tilde{\xi}_{r, \mathbf{k}}
	\right)
	-\frac{\Delta^2}{U} 
	\notag \\
	&\hspace{0.1cm}
	+ 3 U_H \phi^2
	+
	2(\mu_r - \mu_b) \phi
	+ \frac{N}{3V} \left( 2\mu_r + \mu_b - 3\mu_0 \right); \label{gloriginal}
\end{align}
in the expansion in $\phi$ and $\Delta$, we keep in mind that $\mu_r$ and $\mu_b$ are implicit functions of $\Delta$ and $\phi$
through the number equations
\begin{align}
	\phi
	&=
	\frac{1}{V}\sum_\mathbf{k}
	\frac{1}{2}
	\left(
	1
	-
	\xi_{r, \mathbf{k}} \frac{\tanh \beta \tilde{\varepsilon}_k / 2}{\tilde{\varepsilon}_k}
	\right)
	-
	\frac{N}{3V} \label{phi_r}
\end{align}
and
\begin{align}
	-2\phi
	&=
		\frac{1}{V}
	\sum_\mathbf{k}
	\frac{1}{e^{\beta \tilde{\xi}_{b, \mathbf{k}}} + 1}
	-
	\frac{N}{3V}
	\label{phi_b}.
\end{align}
The Ginzburg-Landau free energy up to fourth order in the order parameters is
\begin{align}
	&\mathcal{F}_{GL}(\Delta, \phi)
	\notag \\
	&\hspace{.1cm}=
	a \Delta^2 + \left(b + \frac{(c_2)^2}{c_1} \right)\Delta^4 + 3\left( \frac{1}{c_1} - U_H \right) \phi^2
	\notag \\
	&\hspace{2cm} + c_3 \phi^3 + c_4 \phi^4
	-2\frac{c_2}{c_1}\Delta^2 \phi + c_5 \Delta^2 \phi^2, \label{glthree}
\end{align}
where $c_1 \sim c_5$ and $b$ are all positive, but the sign of $a$ depends on temperature.
The detailed coefficients are given in Appendix A.

The physically realized values of the order parameters minimize the Ginzburg-Landau energy; to leading order in the order parameters,  we then have
\begin{align}
	\frac{\del \mathcal{F}_{GL}}{\del \phi}
	&=
	6\left( \frac{1}{c_1} - U_H \right) \phi - 2\frac{c_2}{c_1}\Delta^2
	=
	0,
	\\
	\frac{\del \mathcal{F}_{GL}}{\del \Delta}
	&=
	2\Delta \left[a + 2\left( b + \frac{(c_2)^2}{c_1} \right) \Delta^2 - 2\frac{c_2}{c_1}\phi \right]
	=
	0.
\end{align}
The first condition implies
\begin{align}
	\phi
	=
	\frac{c_2}{3(1 - c_1 U_H)}\Delta^2,
	\label{phicond}
\end{align}
indicating that if the pairing gap is nonzero, the number imbalance is nonzero, and vice versa.
The second condition, combined with Eq.~(\ref{phicond}), implies
\begin{align}
	\Delta \left[a + 2\left( b + 2 \frac{(c_2)^2}{c_1}-\frac{(c_2)^2}{3c_1 (1 - c_1 U_H)} \right) \Delta^2 \right]
	=
	0.
\end{align}
In addition to the solution $\Delta = 0$,  when $a < 0$ this equation has
a second solution, with lower free energy,
\begin{align}
	\Delta^2
	=
	\frac{|a|}{2(b + {c_2}^2/c_1 - {c_2}^2/(3 c_1 (1 - c_1 U_H)))}.
\end{align}
The transition to fermion pairing is at the temperature at which $a = 0$.

The Ginzburg-Landau free energy of three-component ultracold fermions has certain similarities to the Ginzburg-Landau free energy of dense QCD derived in Refs.~\cite{hatsuda_new_2006,yamamoto_phase_2007,baym_axial_2008},
which makes multicomponent ultracold atoms a promising analog of dense QCD.
The Ginzburg-Landau free energy of dense QCD has the form
\begin{align}
	\Omega_{QCD}(d,\sigma)
	&=
	\frac{\alpha^\prime}{2}|d|^2 + \frac{\beta^\prime}{4}|d|^4 + \frac{a^\prime}{2}\sigma^2 - \frac{c^\prime}{3}\sigma^3 + \frac{b^\prime}{4}\sigma^4
	\notag \\
	&\hspace{1cm}
	 - \gamma^\prime |d|^2 \sigma + \lambda^\prime |d|^2 \sigma^2,
	 \label{LG}
\end{align}
where $d$ is the quark-quark pairing order parameter and $\sigma$ is the chiral symmetry breaking order parameter.
We attach primes to the coefficients to avoid possible confusion with similarly labeled quantities used earlier.
The signs of $\alpha^\prime$ and $a^\prime$ depend on the temperature and the strength of the couplings.
As argued in Refs.~\cite{hatsuda_new_2006,yamamoto_phase_2007,baym_axial_2008}, $\beta^\prime$, $c^\prime$, $\gamma^\prime$, and $\lambda^\prime$ are positive.

With the correspondence between the present system and the dense QCD system, $\Delta \leftrightarrow d$ and $\phi \leftrightarrow \sigma$, we see that
the two Ginzburg-Landau free energies have a similar structure.
 Although the original QCD Lagrangian has a local SU(3) gauge symmetry, the Ginzburg-Landau free energy (\ref{LG}), which does not take the  gluonic degrees of freedom explicitly into account, possesses only global SU(3) symmetry.  To this extent, one can construct an analogy with ultracold atomic fermions.  Similarly, Nambu--Jona-Lasinio models of QCD \cite{hatsuda_kunihiro, abuki_baym_hatsuda_yamamoto} also have only global SU(3) symmetry.
Differences between the QCD free energy and that of ultracold fermions are that the sign of $a^\prime$ becomes negative at low temperature whereas the coefficient of $\phi^2$ is always positive, and in addition
 the coefficients  of $\sigma^3$ and $\phi^3$ are opposite in sign.
These differences are due to the fact that the dense QCD system can undergo chiral symmetry breaking without quark-quark pairing,
but the three-component ultracold fermion system, beginning with equal populations, cannot spontaneously develop local number imbalance without fermion pairing;
with the symmetric interaction we are assuming,
 number imbalance arises from the gain of pairing energy with an increasing number of paired particles.
It would be interesting to see how the analogy can be sharpened in multi-component atomic systems where spontaneous number imbalance and fermion pairing occur independently, for example, with increased numbers of species or with deviations from fully symmetric interactions.

\section{BEC limit}

We turn now to the BEC limit, where the scattering length between fermions is small and positive.  
We can regard the system here as a collection of three types of weakly interacting bound Bose molecules, each made of two fermions, which can be red-green, green-blue, or blue-red. 
The molecules Bose-condense at sufficiently low temperature.  The condensate of molecules can be reduced to a condensate of one type of molecule by appropriately choosing the color axes, as with pairing in the BCS regime.  The condensate in the BEC limit is composed of the same two colors that are paired in the BCS limit.

At high temperature, the system is not condensed, but is simply a gas of thermally excited molecules.
Unlike in the condensate, one cannot exclude the existence of three types of thermally excited molecules.
Whether the high-temperature system develops a number imbalance depends upon the intermolecular interactions.
For the same type of molecules, the effective scattering length is $0.6a$~\cite{petrov_2004},
where $a$ is the scattering length of the constituent fermions.
Between different molecules, as we show later, the effective scattering length is still $0.6a$.  Thus, above the condensation temperature, the system is described by three kinds of thermally excited molecules with the same interaction between all molecules.
As we show in Appendix B, the uncondensed Bose system does not develop a spontaneous number imbalance as long as the interaction between the same types of bosons is greater than half of the interaction between the different bosons.
Thus the present system does not exhibit number imbalance above the condensate transition temperature.

We have, therefore, the following picture of the BEC limit.  At high temperature the system is a homogeneous mixture of three types of molecules.
The Bose-Einstein condensation temperature is that of noninteracting bosons of mass $2m$ and density $N/6V$,
\begin{align}
	T_\mathrm{BEC}
	=
	\frac{\pi}{m[\zeta (3/2)]^{2/3}}\left( \frac{N}{6V} \right)^{2/3}
	\approx
	0.137 \,T_F.
\end{align}
Below $T_\mathrm{BEC}$, the system is a mixture of the condensate of one type of molecule and a cloud of thermal molecules of three types, which vanishes at $T=0$.

We now show that the scattering length between different molecules is the same as that, $0.6a$, between like molecules.  The derivation  of Ref.~\cite{petrov_2004}
 of the scattering length between similar molecules depended on the
 symmetry of the four-particle scattering wave function.  Since, as we show, the wave function for scattering of different molecules has the same symmetry, the arguments of Ref.~\cite{petrov_2004} lead to the same scattering length.
 We write the four-particle scattering wave function between similar molecules, for example, red-green on red-green,
as $\Psi_s (\mathbf{r}_1, \mathbf{r}_2; \mathbf{r}_3, \mathbf{r}_4)$, where $\mathbf{r}_1$ denotes the position of the red fermion of the first molecule,
$\mathbf{r}_2$ is the position of the green fermion of the first molecule,
$\mathbf{r}_3$ is the red fermion of the second molecule, and $\mathbf{r}_4$ is the green fermion of the second molecule.
The symmetries due to Fermi statistics are
\begin{align}
	\Psi_s (\mathbf{r}_1, \mathbf{r}_2; \mathbf{r}_3, \mathbf{r}_4)
	&=
	-\Psi_s (\mathbf{r}_3, \mathbf{r}_2; \mathbf{r}_1, \mathbf{r}_4)
 \notag \\&
	=
	-\Psi_s (\mathbf{r}_1, \mathbf{r}_4; \mathbf{r}_3, \mathbf{r}_2). 
\end{align}

\begin{figure}[htbp]
\begin{center}
\includegraphics[width=6cm,keepaspectratio=true]{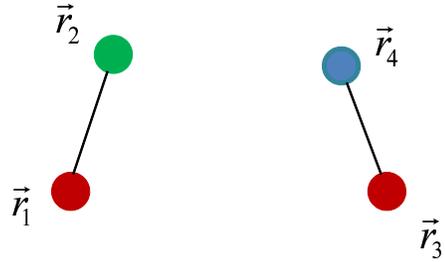}
\end{center}
\caption{(Color online) Two different molecules colliding.}
\label{molecules}
\end{figure}

On the other hand, scattering between different molecules, for example, red-green and red-blue shown in Fig.~\ref{molecules}, described by 
the four-particle scattering wavefunction $\Psi_d (\mathbf{r}_1, \mathbf{r}_2; \mathbf{r}_3, \mathbf{r}_4)$
(where $\mathbf{r}_4$ now denotes the position of the blue fermion),
has only a single symmetry due to Fermi statistics,
\begin{align}
	\Psi_d (\mathbf{r}_1, \mathbf{r}_2; \mathbf{r}_3, \mathbf{r}_4)
	&=
	-\Psi_d (\mathbf{r}_3, \mathbf{r}_2; \mathbf{r}_1, \mathbf{r}_4). \label{symmetry1}
\end{align}
However, for s-wave scattering,
the wave function is symmetric with respect to the interchange of molecules, so that
\begin{align}
	\Psi_d (\mathbf{r}_1, \mathbf{r}_2; \mathbf{r}_3, \mathbf{r}_4)
	&=
	\Psi_d (\mathbf{r}_3, \mathbf{r}_4; \mathbf{r}_1, \mathbf{r}_2). \label{symmetry2}
\end{align}
Conditions (\ref{symmetry1}) and (\ref{symmetry2}) imply that
\begin{align}
	\Psi_d (\mathbf{r}_1, \mathbf{r}_2; \mathbf{r}_3, \mathbf{r}_4)
	&=
	-\Psi_d (\mathbf{r}_1, \mathbf{r}_4; \mathbf{r}_3, \mathbf{r}_2), \label{symmetry3}
\end{align}
which is exactly the same symmetry that was present due to the exchange of green fermions in $\Psi_s$.

The Schr\"odinger equation in the two cases has one apparent difference, that is, the delta-function interaction between the green and blue fermions.
However, the antisymmetry (\ref{symmetry3}) for exchange of green and blue fermions implies that the product of the 
 green-blue potential and the wave function in the Schr\"odinger equation vanishes, so that the Schr\"odinger equation is the same as for identical molecules, and the scattering length is also the same.  This argument depends crucially 
on the two molecules having one color (here red) in common.

\section{Crossover theory}

The crossover, in a two-component system, from  
BCS pairing in the weak-coupling region to a
BEC of weakly interacting molecules in the strong-coupling region is continuous, as seen in experiment ~\cite{greiner_regal_jin_2003, zwierlein_2004, bartenstein_2004} and understood theoretically~\cite{leggettbcsbec,nozieres_schmitt_rink, haussmann_1992, sademelo_1993, pieri_pisani_strinati_2004, hu_drummond_liu_2007, levin_2009, haussmann_1999, pieri_2004}.
A common feature of theories of the BCS-BEC crossover at nonzero temperature is the incorporation of pairing fluctuations,
which allow thermally excited Cooper pairs to exist above the condensate transition temperature.
We now apply this idea to develop
a theory of the crossover, at nonzero temperature, in the three-component system to connect the BCS and BEC regimes discussed earlier, and see that the crossover is also continuous.\footnote{At sufficiently low temperature Efimov states can lower the energy around unitarity, producing a discontinuous transition from the BCS to the BEC regimes ~\cite{floerchinger_functional_2009}.}
We incorporate pairing fluctuations through a self-consistent summation of ladder diagrams, and
then numerically solve for the transition temperature between the condensate and noncondensate phases. 

\subsection{Self-consistent summation of ladder diagrams}

\begin{figure}[htbp]
\begin{center}
\includegraphics[width=8cm,keepaspectratio=true]{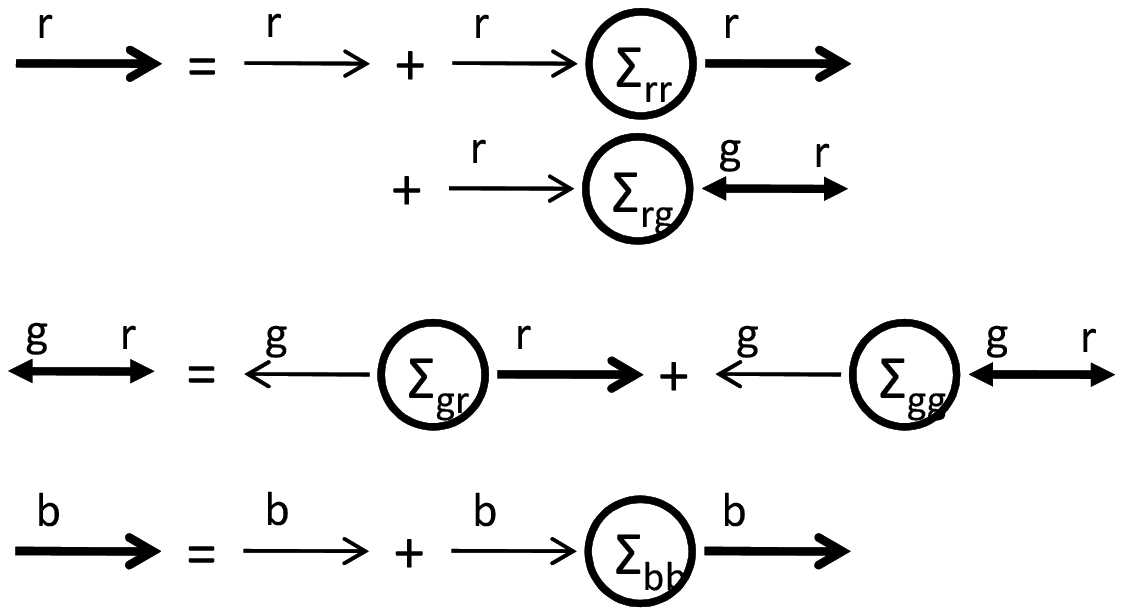}
\end{center}
\caption{The Schwinger-Dyson equations for the normal and anomalous Green's functions.}
\label{dyson}
\end{figure}

\begin{figure}[htbp]
\begin{center}
\includegraphics[width=8cm,keepaspectratio=true]{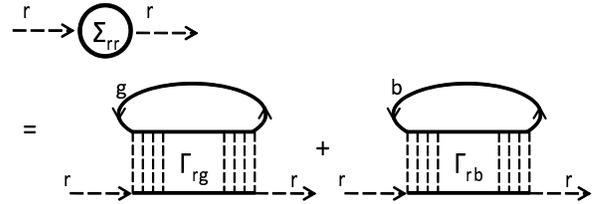}
\end{center}
\caption{Self-energy written in terms of \textit{t} matrices.}
\label{sigma_gamma}
\end{figure}

We construct the crossover theory in terms of the finite temperature
normal and anomalous Green's functions: 
\begin{align}
	\mathcal{G}_\alpha (\mathbf{r - r^\prime}, t-t^\prime)
	&=
	-i\left\langle T\left( \psi_\alpha(\mathbf{r}, t) \psi_\alpha^{\dagger} (\mathbf{r^\prime}, t^\prime) \right) \right\rangle
	\notag\\
	\mathcal{F} (\mathbf{r - r^\prime}, t-t^\prime)
	&=
	-i\left\langle T\left( \psi_r (\mathbf{r}, t) \psi_g (\mathbf{r^\prime}, t^\prime) \right) \right\rangle,
\end{align}
where $T$ denotes time ordering.   We assume still that pairing takes place between r and g particles.
The pairing gap is given in terms of the Fourier transform of $\mathcal{F} (\mathbf{r - r^\prime}, t-t^\prime)$ by
\begin{align}
	\Delta
	&=
	-U\int \frac{d^3 k}{(2\pi)^3}\mathcal{F}(\mathbf{k}, t = 0)
	=
	-\frac{U}{ \beta}\int \frac{d^3 k}{(2\pi)^3}\sum_{\omega_k} \mathcal{F}(k),
\end{align}
where $k$ denotes $(\mathbf{k},\omega_k)$; 
the summation is over the fermionic Matsubara frequencies $\omega_k = i\pi\nu_k/ \beta$ with odd integer $\nu_k$.
The Schwinger-Dyson equations for the Green's functions, illustrated in Fig. \ref{dyson}, are
\begin{align}
	\mathcal{G}_r (k)
	&=
	\mathcal{G}_0 (k)
	+
	\mathcal{G}_0 (k)
	\left(
	\Sigma_{rr} (k)
	\mathcal{G}_r (k)
	+
	\Sigma_{rg} (k) \mathcal{F}^\dagger (k)
	\right), \notag
	\\
	\mathcal{F}^\dagger (k)
	&=
	\mathcal{G}_0 (-k)
	\left(
	-\Sigma_{gr} (k) \mathcal{G}_r (k)
	+
	\Sigma_{gg} (-k) \mathcal{F}^\dagger (k)
	\right), \notag
	\\
	\mathcal{G}_b(k)
	&=
	\mathcal{G}_0 (k)
	+
	\mathcal{G}_0 (k) \Sigma_{bb} (k)
	\mathcal{G}_b(k),
\end{align}
where $\mathcal{G}_0 (k)^{-1} = \omega_k - \xi_\mathbf{k}$ is the free-particle Green's function, and $\Sigma_{\alpha \beta}$ are self-energies with an incoming $\alpha$ particle and an outgoing $\beta$ particle.
Solving this system of equations, we obtain
\begin{align}
	\mathcal{G}_r (k)
	&=
	\left(
	\mathcal{G}_0 (k)^{-1} - \Sigma_{rr} (k) + \frac{\Sigma_{rg}(k)\Sigma_{gr}(k)}{\mathcal{G}_0 (-k)^{-1} - \Sigma_{gg}(-k)}
	\right)^{-1}, \notag
	\\
	\mathcal{G}_b (k)
	&=
	\frac{1}{\mathcal{G}_0 (k)^{-1} - \Sigma_{bb} (k)}, \notag
	\\
	\mathcal{F}^\dagger (k)
	&=
	-\Sigma_{gr} (k)\cdot
	\left\{
	\Sigma_{rg}(k)\Sigma_{gr}(k)
	\right.
	\notag \\
	&\hspace{.1cm}\left.
	+(\mathcal{G}_0(k)^{-1} - \Sigma_{rr}(k))(\mathcal{G}_0(-k)^{-1} - \Sigma_{gg}(-k))
	\right\}^{-1}. \label{fdagger}
\end{align}
The main contribution to the off-diagonal self-energies is the gap:
\begin{align}
 \Sigma_{rg}(k)
	=
	\frac{U}{ \beta} \int \frac{d^3 k^\prime}{(2\pi)^3}\sum_{\omega_{k^\prime}} \mathcal{F}(k^\prime)
	= \Sigma_{gr}(k) =
	-\Delta ,
\end{align}
where we assume without loss of generality that $\Delta$ is real. 
Then the r-particle self-energy, for example, is given by
\begin{align}
	 \Sigma_{rr}(k)
	&=
	-\int \frac{d^3 q}{(2\pi)^3}\frac{1}{ \beta} \sum_{\omega_q}
	\left[
	\Gamma_{rg}(k,k;q) \mathcal{G}_g (-k+q)
	\right.
	\notag \\
	&\hspace{1.5cm}
	\left.
	+ \Gamma_{rb}(k,k;q) \mathcal{G}_b (-k+q)
	\right], \label{sigmarr}
\end{align}
where $\Gamma_{\alpha \beta}(k, k^\prime; q)$, is the two-particle \textit{t} matrix for incoming particles of color $\alpha$ with momenta $k$ and $\beta$ with $-k+q$, and outgoing with momenta $k^\prime$ and $-k^\prime + q$, respectively;
the $\omega_q$ are bosonic Matsubara frequencies. The corresponding diagram is Fig. \ref{sigma_gamma}.
Including the \textit{t} matrix in the self-energy takes pairing fluctuations into account, and as shown in Ref.~\cite{nozieres_schmitt_rink}, encompasses thermal fluctuations of paired molecules in the BEC limit and the Hartree approximation in the BCS limit~\cite{nozieres_schmitt_rink}, thus connecting both limits continuously.
Note that there is no process of this form in which the top line is anomalous
since such a process would involve scattering between two r particles,  either initially or finally, which is forbidden by the Pauli principle; the internal lines can, however, be anomalous.  

On the other hand, in the self-energy of b particles, the top line can in principle be anomalous; however, this process would involve particle-hole scatterings either initially or finally, which is negligible for short-range interactions~\cite{fetterwalecka}; the self-energy involves only a sum of rb and gb particle-particle scatterings.  The Bethe-Salpeter equation for the rb \textit{t} matrix becomes
 \begin{align}
	&\Gamma_{r b}(k, k^\prime; q)
	\notag \\
	&\hspace{.1cm}
	=
	-U
	-U
	\int \frac{d^3 p}{(2\pi)^3}\frac{1}{ \beta} \sum_{\omega_p} \mathcal{G}_r (p) \mathcal{G}_b (-p+q) \Gamma_{r b}(p,k^\prime;q).
	\label{hans}
\end{align}
As one sees by iterating this equation,  $\Gamma_{rb}(k,k^\prime;q)$ is independent of $k$ and $k^\prime$; we write $ \Gamma(k,k^\prime; q)=\Gamma(q)$.
Solving Eq.~(\ref{hans}), we obtain
\begin{align}
	\Gamma_{rb}(q)
	=
	-\left(
	\frac{1}{U} + \int \frac{d^3 p}{(2\pi)^3}\frac{1}{ \beta}\sum_{\omega_p} \mathcal{G}_r (p) \mathcal{G}_{b}(-p+q)
	\right)^{-1};
	 \label{gammarb}
\end{align}
$\Gamma_{gb}$ takes the same form {\it mutatis mutandis}.

\begin{figure}[htbp]
\begin{center}
\includegraphics[width=5cm,keepaspectratio=true]{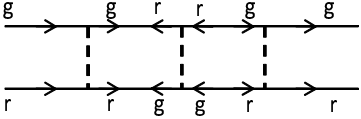}
\end{center}
\caption{An anomalous contribution to the rg \textit{t} matrix.}
\label{rganomalous}
\end{figure}

In $\Gamma_{rg}$ we must take the rg anomalous Green's functions into account, as illustrated in Fig.~\ref{rganomalous}.
Solving the Bethe-Salpeter equation in Nambu matrix notation, we have
\begin{align}
	\Gamma_{rg}(q)
	=
	\frac{\chi_{11}(-q)}{\chi_{11}(q) \chi_{11}(-q) - \chi_{12}(q)^2}, \label{gammarg}
\end{align}
where
\begin{align}
	\chi_{11}(q)
	&=
	-\frac{1}{U} - \int \frac{d^3 p}{(2\pi)^3}\frac{1}{ \beta}\sum_{\omega_p} \mathcal{G}_{r}(p)\mathcal{G}_g (q-p),
	\\
	\chi_{12}(q)
	&=
	\int \frac{d^3 p}{(2\pi)^3}\frac{1}{ \beta}\sum_{\omega_p} \mathcal{F}(p)\mathcal{F}^\dagger (q-p).
\end{align}

To determine the gap and the number imbalance as a function of temperature and scattering length involves self-consistently solving the gap equation (\ref{fdagger}), which can be rewritten as
\begin{align}
	-&\frac{1}{U}
	=
	\int \frac{d^3 k}{(2\pi)^3}\frac{1}{ \beta}\sum_{\omega_k}
	\notag \\
	&\cdot
	\frac{1}{(\mathcal{G}_0 (k)^{-1} - \Sigma_{rr}(k))(\mathcal{G}_0 (-k)^{-1} - \Sigma_{gg}(-k)) + \Delta^2},
\end{align}
together with the number equations
\begin{align}
	\frac{N_r}{V}
	=
	\lim_{\eta \to +0}\int \frac{d^3 k}{(2\pi)^3}\frac{1}{\beta}\sum_{\omega_k} e^{i\omega_k \eta}\mathcal{G}_r (k),
	\\
	\frac{N_b}{V}
	=
	\lim_{\eta \to +0}\int \frac{d^3 k}{(2\pi)^3}\frac{1}{ \beta}\sum_{\omega_k} e^{i\omega_k \eta}\mathcal{G}_b (k).
\end{align}
However in this paper we focus only on calculating the transition temperature.
  
\subsection{Evaluation of $T_c$}
 
 We now use the formalism of the previous subsection to evaluate the transition temperature, where the pairing gap $\Delta$ becomes zero.
The gap equation at $T_c$ is equivalent to the condition that $\Gamma_{rg}(q)$ diverges at $q = 0$.
Therefore, at $T_c$, we can make the approximations,
\begin{align}
	&\Sigma_{rr}(k)
	\notag \\
	&=
	-\int \frac{d^3 q}{(2\pi)^3}\frac{1}{\beta_c}\sum_{\omega_q}
	\left(
	\Gamma_{rg}(q)\mathcal{G}_g(q-k)
	+ \Gamma_{rb}(q) \mathcal{G}_b (q-k)
	\right)
	\notag \\
	&\approx
	-\int \frac{d^3 q}{(2\pi)^3}\frac{1}{\beta_c}\sum_{\omega_q}
	\left(
	\Gamma_{rg}(q)\mathcal{G}_g(-k)
	+ \Gamma_{rb}(q) \mathcal{G}_b (-k)
	\right),\label{sigrr}
\end{align}
and
\begin{align}
	\Sigma_{bb}(k)
	=&
	-\int \frac{d^3 q}{(2\pi)^3}\frac{2}{\beta_c}\sum_{\omega_q}
	\Gamma_{br}(q)\mathcal{G}_r(q-k)
	\notag \\
	&\approx
	-\int \frac{d^3 q}{(2\pi)^3}\frac{2}{\beta_c}\sum_{\omega_q}
	\Gamma_{br}(q) \mathcal{G}_r(-k).
\end{align}
For $T\ge T_c$,  the t-matrices do not depend on the color indices. 
Then, using the final line of Eq.~(\ref{sigrr}) we see that the Green's function for r particles becomes
\begin{align}
	\mathcal{G}_r (k)
	&=
	\left( \mathcal{G}_0^{-1} (k) - \Sigma_{rr}(k) \right)^{-1}
	\notag \\
	&\approx
	\left( \mathcal{G}_0^{-1} (k) + \mathcal{G}_0 (-k) \Delta_{pg}^2 \right)^{-1}
	\notag \\
	&=
	-\frac{\omega_k + \xi_\mathbf{k}}{|\omega_k|^2 + \xi_\mathbf{k}^2 + \Delta_{pg}^2}, \label{grpseudo}
\end{align}
where we introduce a ``pseudogap" $\Delta_{pg}$ at $T_c$ by writing
\begin{align}
	\Delta_{pg}^2
	&=
	\frac{2}{\beta_c}\int \frac{d^3 q}{(2\pi)^3}\sum_{\omega_q} \Gamma(q),
\end{align}
with $\Gamma = \Gamma_{rg} = \Gamma_{rb} = \Gamma_{br}$.
The final line of Eq.~(\ref{grpseudo}) is just a BCS Green's function with the gap replaced by the pseudogap.
We write $E_\mathbf{k} = \sqrt{\xi_\mathbf{k}^2 + \Delta_{pg}^2}$ for convenience.
Similarly $\mathcal{G}_b (k)$ has the same form at $T = T_c$.

The number equations then reduce to
\begin{align}
	\frac{N}{3V}
	&=
	\frac{1}{2}
	\int \frac{d^3 k}{(2\pi)^3}
	\left(
	1
	-
	\frac{\xi_\mathbf{k}}{E_\mathbf{k}}\tanh \frac{\beta_c E_\mathbf{k}}{2}
	\right), \label{nrfinite}
\end{align}
while the equation for the pseudogap is
\begin{align}
	&-\frac{1}{U}
	=
	\int\frac{d^3 k}{(2\pi)^3}\frac{1}{\beta_c} \sum_{\omega_k} \mathcal{G}_r (k) \mathcal{G}_g (-k)
	\notag \\
	&=
	\int \frac{d^3 k}{(2\pi)^3}\left\{\left( 1 + \frac{\xi_\mathbf{k}^2}{E_\mathbf{k}^2} \right) \frac{\tanh (\beta_c E_\mathbf{k}/2)}{4E_\mathbf{k}}
	-\frac{\Delta_{pg}^2}{E_\mathbf{k}^2}\frac{f^\prime (E_\mathbf{k})}{2}
	\right\};
	\label{tceq}
\end{align}
as before, the bare coupling $U$ is related to the scattering length $a$ through Eq.~(\ref{renormalizeu}).

In the BCS limit, $k_Fa \to 0^-$, $\Delta_{pg}^2$ tends to zero, as we can see by considering 
the BCS gap equation at $T_c$ (not the mean-field BCS transition
temperature, but the same $T_c$ that we are using here) with a gap
$\Delta$
\begin{align}
       &-\frac{1}{U}
       =
       \int \frac{d^3 k}{(2\pi)^3}\left\{ \frac{\tanh (\beta_c
\sqrt{\xi_\mathbf{k}^2 + \Delta^2}/2)}{2\sqrt{\xi_\mathbf{k}^2 +
\Delta^2}}
       \right\}. \label{tcbcseq}
\end{align}
Expanding the right sides of (\ref{tceq}) and (\ref{tcbcseq}) in
terms of $\Delta_{pg}^2$ and $\Delta^2$, we see that the zeroth
order terms are identical.
Also, since the final line of Eq.~(\ref{tceq}) decreases monotonically with $\Delta_{pg}^2$, the limit
$\Delta^2 \to 0$, as in weak-coupling BCS, implies $\Delta_{pg}^2 \to 0$.  

Determining $T_c$ requires estimating $\Delta_{pg}^2$, which we do by
 expanding $\Gamma_{rg} (q)^{-1}$ around $q = 0$, recalling that  $\Gamma_{rg}(0)^{-1} = 0$ at $T_c$:
\begin{align}
	&-\Gamma_{rg} (\mathbf{q}, \omega_q)^{-1}
	=
	\frac{1}{U}
	+
	\int \frac{d^3 p}{(2\pi)^3} \frac{1}{\beta_c}\sum_{\omega_p}\mathcal{G}_r (p)\mathcal{G}_g (q-p)
	\notag \\
	&\hspace{1cm}
	\approx
	\int \frac{d^3 p}{(2\pi)^3} \frac{1}{\beta_c}\sum_{\omega_p}\mathcal{G}_r (p)
	\notag \\
	&\hspace{1cm}
	\cdot
	\left\{
	\left.\frac{\partial}{\partial \omega}\mathcal{G}_g (\mathbf{k},\omega) \right|_{k = -p}\omega_q
	+ \frac{1}{6}\left.\nabla^2 \mathcal{G}_g (\mathbf{k},\omega) \right|_{k = -p}q^2
	\right\}
	\notag \\
	&\hspace{1cm}\equiv
	Z \omega_q - \gamma q^2. \label{gammaexpand}
\end{align}
Explicit forms for $Z$ and $\gamma$ are given in Appendix C.
The pseudogap then becomes
\begin{align}
	\Delta_{pg}^2
	&=
	-2\int \frac{d^3 q}{(2\pi)^3}\frac{1}{\beta_c} \sum_{\omega_q} \frac{1}{Z\omega_q - \gamma q^2}
	\notag \\
	&=
	2\frac{1}{Z}\int \frac{d^3 q}{(2\pi)^3}\frac{1}{e^{\beta_c \gamma q^2 / Z} - 1}
	=
	\frac{\zeta (3/2)}{4Z}\left( \frac{Z}{\pi \beta_c \gamma}\right)^{3/2}.
	 \label{pseudogapexp}
\end{align}

\begin{figure}[htbp]
\begin{center}
\includegraphics[width=9cm,keepaspectratio=true]{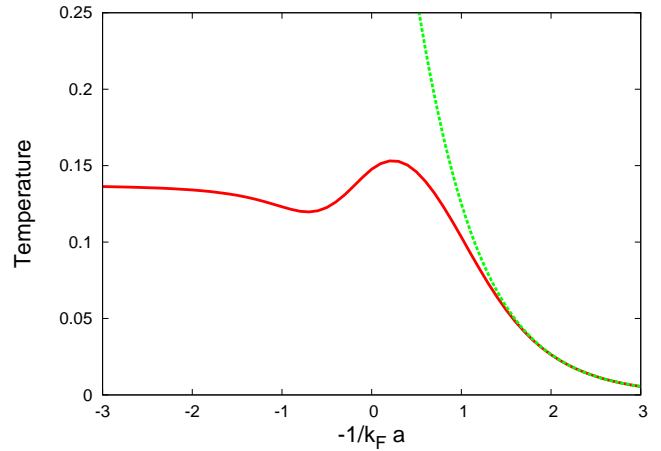}
\end{center}
\caption{(Color online) The phase diagram of three-component ultracold Fermi gas.  The temperature is in units of $T_F$. The solid line is the transition temperature calculated with pairing fluctuations incorporated through the summation of ladder diagrams. The dotted line is the transition temperature calculated from mean-field BCS theory. The mean-field line corresponds roughly to the temperature at which fermions start to form (noncondensed) pairs. The line calculated from the ladder summation is where the Cooper pairs start to condense. Toward the left end of the figure, the transition temperature approaches the BEC limiting value $T_c \sim 0.137\, T_F$.}
\label{tc}
\end{figure}

Solving the number equation (\ref{nrfinite}), the gap equation (\ref{tceq}), and the expression for the pseudogap (\ref{pseudogapexp})
self-consistently, we obtain the transition temperature, plotted 
against $-1/k_F a$ in Fig. \ref{tc}.
The solid line in the figure is the transition temperature calculated with the ladder summation formalism described here,
and the dotted line is the result from mean-field BCS theory.
The ladder summation line approaches the mean-field line in the BCS limit.
On the other hand, in the BEC limit, the ladder summation correctly yields $T_c \to 0.137\, T_F$.
The crossover theory presented here connects both limits continuously.

Throughout, we have kept a common chemical potential for the different species, and found that below $T_c$ the number of b particles becomes smaller than the number of r or g particles.
In ultracold atomic experiments, the number of the particles in each species is usually fixed at the start, and thus
the simplest scenario that may occur experimentally is that the number imbalance appears through the formation of population-imbalanced domains. 
The formation of population-imbalanced domains leads to a gain of condensation energy of order $E_cV/2$ for the fully imbalanced state, where $E_c$ is the condensate energy density in a balanced system; the factor $1/2 = 3/2 - 1$ is the increase in the relative number of Cooper pairs in the imbalanced state over that in the balanced state.
On the other hand, the formation of a single domain wall costs a net surface energy $E_{surf}$ of order $E_cV \xi_c /L$,  where $\xi_c$ is the coherence length and $L$ is the linear size of the system.
The condition that the formation of the domain is beneficial for the system is $E_cV / 2 > E_{surf}$, or roughly 
$L/\xi_c \simge 1$, which typically holds well.
Domain formation is expected to decrease the free energy from that of the homogeneous state at low temperature.
Other possible realizations of population imbalance include the formation of a ``color density" wave or the formation of an inhomogeneous (Fulde-Ferrell-Larkin-Ovchinnikov) superfluid; we leave analysis of these states as future study.   Also to apply the present theory quantitatively under realistic experimental conditions it will be necessary to investigate the effects of Efimov states.

\begin{acknowledgements}
This work was supported in part by the National Science Foundation under Grants No.~PHY07-01611 and No.~PHY09-69790.
T.O. would like to thank Kenji Maeda for helpful comments and Jussi Kajala for his help with the numerical calculations.
\end{acknowledgements}

\appendix
\section{The coefficients of the Ginzburg-Landau free energy}

We outline here the derivation of the Ginzburg-Landau free energy (\ref{glthree}) from the free energy $\mathcal{F}_{\mathrm{GL}}(\Delta, \phi)$, Eq.~(\ref{gloriginal}).
Since $\Delta$ always appears squared in the equations, odd powers of $\Delta$ do not occur in the free energy.
To find the coefficients of $\Delta^2$ and $\Delta^4$, we set $\phi = 0$,
and expand $\mathcal{F}_{\mathrm{GL}}(\Delta, 0)$ in powers of $\Delta^2$.
Taking the derivative of the number equation for blue particles (\ref{phi_b}) with respect to $\Delta^2$,
we see that $\mu_b$ (here allowed to differ from $\mu_r$) does not depend on $\Delta^2$.  Differentiating the number equation for red particles (\ref{phi_r}), we obtain
\begin{align}
	\left. \frac{\del \mu_r}{\del \Delta^2} \right|_{0} = -\frac{c_2}{c_1},
\end{align}
where the subscript 0 denotes the derivative at $\Delta = \phi = 0$, and
\begin{align}
	c_1
	&=
	-\frac{1}{V}\sum_\mathbf{k} f^\prime (\xi_\mathbf{k}^0),
	\\
	c_2
	&=
	\frac{1}{V}\sum_\mathbf{k} \left( \frac{\tanh \beta \xi_\mathbf{k}^0/2}{4 (\xi_\mathbf{k}^0)^2} + \frac{f^\prime (\xi_\mathbf{k}^0)}{2\xi_\mathbf{k}^0}\right).
\end{align}
Note that both $c_1$ and $c_2$ are positive.
Then
\begin{align}
	 \left.\frac{\del \mathcal{F}_{\mathrm{GL}}(\Delta,0)}{\del \Delta^2}\right|_0
	 &=
	 -\frac{1}{U} - \frac{1}{V}\sum_\mathbf{k} \frac{\tanh \beta \xi_\mathbf{k}^0/2}{2\xi_\mathbf{k}^0},
	 \\
	 \frac{1}{2}\left.\frac{\del^2 \mathcal{F}_{\mathrm{GL}}(\Delta,0)}{\del (\Delta^2)^2}\right|_0
	 &=
	 b + \frac{(c_2)^2}{c_1},
\end{align}
where
\begin{align}
	b
	=
	\frac{1}{V}\sum_\mathbf{k}
	\left(
	\frac{\tanh \beta \xi_\mathbf{k}^0/2}{8(\xi_\mathbf{k}^0)^3}
	+
	\frac{f^\prime (\xi_\mathbf{k}^0)}{4(\xi_\mathbf{k}^0)^2}
	\right) >0.
\end{align}

We similarly derive the coefficients of $\phi$, $\phi^2$, $\phi^3$, and $\phi^4$:
\begin{align}
	\left. \frac{\del \mathcal{F}_{\mathrm{GL}}(0,\phi)}{\del \phi} \right|_{0}
	&=
	0,
	\\
	\frac{1}{2}\left. \frac{\del^2 \mathcal{F}_{\mathrm{GL}}(0,\phi)}{\del \phi^2} \right|_{0}
	&=
	3\left( \frac{1}{c_1} - U_H \right),
	\\
	\frac{1}{6}\left. \frac{\del^3 \mathcal{F}_{\mathrm{GL}}(0,\phi)}{\del \phi^3} \right|_{0}
	&=
	\frac{\kappa_1}{(c_1)^3},
	\\
	\frac{1}{24}\left. \frac{\del^4 \mathcal{F}_{\mathrm{GL}}(0,\phi)}{\del \phi^4} \right|_{0}
	&=
	\frac{3}{4 (c_1)^4} \left( 3 \frac{(\kappa_1)^2}{c_1} - \kappa_2 \right),
\end{align}
where
\begin{align}
	\kappa_1
	&=
	\frac{1}{V}\sum_\mathbf{k} f^{\prime \prime}(\xi_\mathbf{k}^0),
	&
	\kappa_2
	&=
	-\frac{1}{V}\sum_\mathbf{k} f^{\prime \prime \prime}(\xi_\mathbf{k}^0).
\end{align}

Finally, the coefficients of $\phi \Delta^2$ and $\phi^2 \Delta^2$  are
\begin{align}
	&  \left. \frac{\del^2 \mathcal{F}_{\mathrm{GL}}(\Delta, \phi)}{\del\phi\del \Delta^2} \right|_{\Delta = \phi = 0}
	=
	-2\frac{c_2}{c_1},
	\\
	&\frac{1}{2}\left. \frac{\del^3 \mathcal{F}_{\mathrm{GL}}(\Delta, \phi)}{(\del \phi)^2 \del \Delta^2} \right|_{\Delta = \phi = 0}
	\notag \\
	&\hspace{0.1cm}=
	\frac{c_2 \kappa_1}{(c_1)^3}
	+
	\frac{1}{(c_1)^2}
	\left(
	\frac{1}{V}\sum_\mathbf{k} \frac{f^{\prime \prime}(\xi_\mathbf{k}^0)}{2\xi_\mathbf{k}^0}
	-
	4b
	\right)
	\equiv
	c_5.
\end{align}
Therefore, the Ginzburg-Landau free energy up to fourth order in the order parameters is
\begin{align}
	&\mathcal{F}_{\mathrm{GL}}(\Delta, \phi)
	\notag \\
	&\hspace{0.1cm}=
	\left( -\frac{1}{U} - \frac{1}{V}\sum_\mathbf{k} \frac{\tanh \beta \xi_\mathbf{k}^0/2}{2\xi_\mathbf{k}^0} \right) \Delta^2
	+
	\left( b + \frac{(c_2)^2}{c_1} \right) \Delta^4
	\notag \\
	&\hspace{0.2cm}
	+
	3\left( \frac{1}{c_1} - U_H \right) \phi^2
	+
	\frac{\kappa_1}{(c_1)^3} \phi^3
	\notag \\
	&\hspace{0.2cm}
	+
	\frac{3}{4 (c_1)^4} \left( 3 \frac{(\kappa_1)^2}{c_1} - \kappa_2 \right) \phi^4
	-
	2\frac{c_2}{c_1} \phi \Delta^2
	+
	c_5 \phi^2 \Delta^2
	\notag \\
	&\hspace{0.1cm}\equiv
	a \Delta^2 + \left(b + \frac{(c_2)^2}{c_1} \right)\Delta^4 + 3\left( \frac{1}{c_1} - U_H \right) \phi^2
	\notag \\
	&\hspace{0.2cm} + c_3 \phi^3 + c_4 \phi^4
	-2\frac{c_2}{c_1}\Delta^2 \phi + c_5 \Delta^2 \phi^2.
\end{align}
Note that the $c_i$ and $b$ are all positive.
Also, since $U_H$ is negative, the coefficient of $\phi^2$ is positive.

\section{Population imbalance in a Bose mixture above the condensation temperature}

We derive the condition for the homogeneous state with population balance to be stable. Although the three-component ultracold Fermi gas can form three types of molecules,
the basic physics of the instability toward inhomogeneous states can be captured by considering a two-component Bose system.

We derive the Ginzburg-Landau free energy of a system of two species of bosons, a and b, as a function of their  population imbalance at fixed total number $N= N_a+N_b$.   With $a_\mathbf{k}$ and $b_\mathbf{k}$ the annihilation operators of bosons $a$ and $b$ of momentum $\mathbf{k}$,
the Hamiltonian is
\begin{align}
	&H-\mu_a N_a - \mu_b N_b
	\notag \\
	&\hspace{0.1cm}
	=
	\sum_\mathbf{k}
	\left(
	\frac{ k^2}{2m} - \mu_a
	\right)
	a_\mathbf{k}^\dagger a_\mathbf{k}
	+
	\sum_\mathbf{k}
	\left(
	\frac{ k^2}{2m} - \mu_b
	\right)
	b_\mathbf{k}^\dagger b_\mathbf{k}
	\notag \\
	&\hspace{1cm}
	+
	\frac{U_0}{2V}\sum_{\mathbf{k},\mathbf{k}^\prime,\mathbf{q}}
	\left(
	a^\dagger_\mathbf{k+q} a^\dagger_\mathbf{k^\prime - q} a_\mathbf{k^\prime} a_\mathbf{k}
	+
	b^\dagger_\mathbf{k+q} b^\dagger_\mathbf{k^\prime - q} b_\mathbf{k^\prime} b_\mathbf{k}
	\right)
	\notag \\
	&\hspace{1cm}+
	\frac{U_1}{V}
	\sum_{\mathbf{k},\mathbf{k}^\prime,\mathbf{q}}
	a^\dagger_\mathbf{k+q} b^\dagger_\mathbf{k^\prime - q} b_\mathbf{k^\prime} a_\mathbf{k},
\end{align}
where $U_0 = 4\pi a_0/m$ and $U_1 = 4\pi a_1/m$ are the s-wave interaction strength between the same type and between different types of bosons,
and $a_0$ and $a_1$ are the corresponding scattering lengths.

We assume a sufficiently high temperature that neither system is condensed.  In the Hartree-Fock approximation, we obtain
\begin{align}
	&H-\mu_a N_a - \mu_b N_b
	\notag \\
	&\hspace{0.1cm}\approx
	\sum_\mathbf{k}
	\left(
	\xi_{a, \mathbf{k}} + 2U_0\frac{N_a}{V} + U_1 \frac{N_b}{V}
	\right)
	a_\mathbf{k}^\dagger a_\mathbf{k}
	\notag \\
	&\hspace{0.2cm}+
	\sum_\mathbf{k}
	\left(
	\xi_{b, \mathbf{k}} + 2U_0\frac{N_b}{V} + U_1 \frac{N_a}{V}
	\right)
	b_\mathbf{k}^\dagger b_\mathbf{k}
	\notag \\
	&\hspace{0.2cm}
	-
	\frac{U_0}{V}\left( N_a^2 +N_b^2 \right)
	-
	\frac{U_1}{V}N_a N_b,
\end{align}
where  $\xi_{a, \mathbf{k}} \equiv k^2/2m - \mu_a$ and $\xi_{b, \mathbf{k}} \equiv k^2/2m - \mu_b$.
The number of particles $N_a$ and $N_b$ satisfies the self-consistent equations,
\begin{align}
	N_a
	&=
	\sum_\mathbf{k}
	g\left(
	\xi_{a, \mathbf{k}} + 2U_0n_a + U_1 n_b
	\right) \label{numbera}
	\\
	N_b
	&=
	\sum_\mathbf{k}
	g\left(
	\xi_{b, \mathbf{k}} + 2U_0n_b + U_1 n_a
	\right), \label{numberb}
\end{align}
where $g(x) = 1/(e^{\beta x} - 1)$ is the Bose distribution function, and
 $n_a = N_a / V$ and $n_b = N_b / V$.
Then the thermodynamic potential is
\begin{align}
	\frac{\Omega}{V}
	&=
	-U_0 \left( n_a^2 +n_b^2 \right) - U_1 n_a n_b
	\notag \\
	&\hspace{0.1cm}+
	\frac{1}{\beta V}
	\sum_\mathbf{k}
	\ln \left\{ 1 - \exp \left(-\beta \left( \xi_{a, \mathbf{k}} + 2U_0 n_a + U_1 n_b \right) \right)\right\}
	\notag \\
	&\hspace{0.1cm}
	+
	\frac{1}{\beta V}
	\sum_\mathbf{k}
	\ln \left\{ 1 - \exp \left(-\beta \left( \xi_{b, \mathbf{k}} + 2U_0 n_b + U_1 n_a \right) \right)\right\},
\end{align}
and the Helmholtz free energy is
\begin{align}
	\frac{F}{V}
	&=
	 \frac{\Omega}{V} + \mu_a n_a + \mu_b n_b.
\end{align} 

The condition for the stability of the homogeneous state is found by expanding the 
Helmholtz free energy in terms of the deviation of the number of particles from the homogeneous state.
We write the deviation of the numbers of particles from the balanced case as
\begin{align}
	\varphi = n_a - \frac{n}{2} = -\left(n_b- \frac{n}{2}\right) ;
\end{align}
then 
\begin{align}
	\left. \frac{\del}{\del \varphi} \frac{F}{V}\right|_{\varphi = 0} &= 0
	\\
	\left.\frac{\del^2}{\del \varphi^2}\frac{F}{V}\right|_{\varphi = 0}
	&=
	2\left( 2U_0 - U_1 - \frac{1}{G}\right),\label{coef_phi2}
\end{align}
where
\begin{align}
	G
	=
	\frac{1}{V}
	\sum_\mathbf{k}
	g^\prime \left( \frac{ k^2}{2m} - \mu_0 + 2U_0 \frac{n}{2} + U_1\frac{n}{2} \right) <0,
\end{align}
and the homogeneous chemical potential $\mu_0$ is determined by
\begin{align}
	\frac{n}{2}
	=
	\frac{1}{V}\sum_\mathbf{k}
	g\left( \frac{ k^2}{2m} - \mu_0 + 2U_0 \frac{n}{2} + U_1  \frac{n}{2} \right).
	\end{align}
The homogeneous state is stable if and only if $\del^2(F/V)/\del\varphi^2>0$.  Since $G<0$, we immediately conclude that when $2U_0 > U_1$, as in the present system, the homogeneous state is always stable at $T > T_\mathrm{BEC}$.
[For $2U_0 < U_1$, one finds $G \to 0^-$ as $T \to \infty$, and $G \to -\infty$ as $T$ approaches  $T_\mathrm{BEC}$ from above, implying
 a phase transition from the homogeneous to an inhomogeneous state at $T>T_\mathrm{BEC}$.
The transition temperature increases with increasing $U_1 - 2U_0$.
As  $U_1 \to 2U_0 $ from above,  the transition temperature approaches $T_\mathrm{BEC}$ from above.]

Since the interaction is the same as that between identical and different molecules in the BEC limit of three-component ultracold fermions,
the result derived here implies that the system is homogeneous above the condensation temperature.\\

\section{Expansion of $\Gamma_{rg}(\mathbf{q}, \omega_q)^{-1}$}

The expansion of $\Gamma_{rg}(\mathbf{q}, \omega_q)^{-1}$  can be explicitly carried out
using Eq.~(\ref{grpseudo}), with the result of Eq.~(\ref{gammaexpand}), $-\Gamma_{rg}(\mathbf{q}, \omega_q)^{-1} \approx Z \omega_q - \gamma q^2$, where
\begin{align}
	Z
	&=
	\int \frac{d^3 k}{(2\pi)^3}
	\left(
	\frac{\tanh (\beta_c E_\mathbf{k}/2)}{2E_\mathbf{k}} + f^\prime (E_\mathbf{k})
	\right)
	\frac{\xi_\mathbf{k}}{2E_\mathbf{k}^2}
\end{align}
and
\begin{align}
	\gamma
	&=
	\int \frac{d^3 k}{(2\pi)^3}\frac{1}{2mE_\mathbf{k}^7}
	\left[
	\left\{
	\xi_\mathbf{k}^2 \Delta_{pg}^2 \frac{ k^2}{3m}
	+
	\frac{1}{8}\xi_\mathbf{k} E_\mathbf{k}^2 \left( 2\xi_\mathbf{k}^2 - \Delta_{pg}^2\right)
	\right. \right.
	\notag \\
	&\hspace{0.1cm}
	\left. \left.
	+\frac{ k^2}{24m}\left(\Delta_{pg}^2 - \xi_\mathbf{k}^2\right)
		\left( E_\mathbf{k}^2 + \xi_\mathbf{k}^2 \right)
	\right\}
	\left(
	\tanh \frac{\beta_c E_\mathbf{k}}{2} + 2E_\mathbf{k} f^\prime (E_\mathbf{k})
	\right)
	\right.
	\notag \\
	&\hspace{0.1cm}
	+
	\left.
	\left\{
	\frac{\xi_\mathbf{k} \Delta_{pg}^2 E_\mathbf{k}^2}{4}
	+
	\frac{k^2}{12m}
	\left(
	2\xi_\mathbf{k}^4 - \xi_\mathbf{k}^2 \Delta_{pg}^2 + \Delta_{pg}^4
	\right)
	\right\}
	E_\mathbf{k}^2 f^{\prime \prime}(E_\mathbf{k})
	\right.
	\notag \\
	&\hspace{0.1cm}
	+
	\left.
	\frac{k^2}{18m} \xi_\mathbf{k}^2 \Delta_{pg}^2 E_\mathbf{k}^3 f^{\prime \prime \prime}(E_\mathbf{k})
	\right].
\end{align}

\end{document}